\pdfoutput=1
\documentclass[sigconf]{acmart}
\usepackage{balance}
\usepackage{enumitem}
\usepackage{multirow}
\usepackage{booktabs}
\usepackage{graphicx}
\AtBeginDocument{%
  }

\copyrightyear{2025}
\acmYear{2025}
\setcopyright{acmlicensed}
\acmConference[SIGIR '25] {Proceedings of the 48th International ACM SIGIR Conference on Research and Development in Information Retrieval}{ July 13--18, 2025}{Padua, Italy.}
\acmBooktitle{Proceedings of the 48th International ACM SIGIR Conference on Research and Development in Information Retrieval (SIGIR '25), July 13--18, 2025, Padua, Italy}
\acmISBN{979-8-4007-1592-1/25/07}
\acmDOI{10.1145/3726302.3730228}

\usepackage{xcolor}
\usepackage{soul}
\usepackage{comment}
\usepackage{amssymb}

\setuldepth{cat}

\newcommand{\note}[1]{{\color{magenta}$\square$}}

\begin{document}

\title{LREA: Low-Rank Efficient Attention on Modeling Long-Term User Behaviors for CTR Prediction }

\author{Xin Song}
\affiliation{
  \institution{Alibaba International Digital Commerce Group}
  \city{Beijing} 
  \state{} 
  \country{China}
}
\email{songxin.song@alibaba-inc.com}

\author{Xiaochen Li}
\affiliation{
  \institution{Alibaba International Digital Commerce Group}
  \city{Beijing} 
  \state{} 
  \country{China}
}
\email{xingke.lxc@alibaba-inc.com}

\author{Jinxin Hu}\authornote{Corresponding author.}
\affiliation{
  \institution{Alibaba International Digital Commerce Group}
  \city{Beijing} 
  \state{} 
  \country{China}
}
\email{jinxin.hjx@alibaba-inc.com}

\author{Hong Wen}
\affiliation{
  \institution{Unaffiliated}
  \city{} 
  \state{} 
  \country{}
}
\email{dreamonewh@gmail.com}

\author{Zulong Chen}
\affiliation{
  \institution{Alibaba Group}
  \city{Beijing} 
  \state{} 
  \country{China}
}
\email{zulong.czl@alibaba-inc.com}

\author{Yu Zhang}
\affiliation{
  \institution{Alibaba International Digital Commerce Group}
  \city{Beijing} 
  \state{} 
  \country{China}
}
\email{daoji@alibaba-inc.com}

\author{Xiaoyi Zeng}
\affiliation{
  \institution{Alibaba International Digital Commerce Group}
  \city{Hangzhou} 
  \state{} 
  \country{China}
}
\email{yuanhan@alibaba-inc.com}

\author{Jing Zhang}
\authornotemark[1]
\affiliation{
  \institution{School of Computer Science, Wuhan University}
  \city{Wuhan} 
  \state{} 
  \country{China}
}
\email{jingzhang.cv@gmail.com}


\renewcommand{\shortauthors}{Xin Song et al.}

\begin{abstract}
With the rapid growth of user historical behavior data, user interest modeling has become a prominent aspect in Click-Through Rate (CTR) prediction, focusing on learning user intent representations. However, this complexity poses computational challenges, requiring a balance between model performance and acceptable response times for online services. 
Traditional methods often utilize filtering techniques. These techniques can lead to the loss of significant information by prioritizing top K items based on item attributes or employing low-precision attention mechanisms. 
In this study, we introduce LREA, a novel attention mechanism that overcomes the limitations of existing approaches while ensuring computational efficiency. LREA leverages low-rank matrix decomposition to optimize runtime performance and incorporates a specially designed loss function to maintain attention capabilities while preserving information integrity. 
During the inference phase, matrix absorption and pre-storage strategies are employed to effectively meet runtime constraints. 
The results of extensive offline and online experiments demonstrate that our method outperforms state-of-the-art approaches.
\end{abstract}




\maketitle



\section{Introduction}

Click-Through Rate (CTR) prediction is pivotal in online advertising and recommendation systems. Previous studies on short sequences have mainly focused on the interaction between sequences and targets or trigger item~\cite{DIN,DIEN,UID,DIHN}. With the rapid growth of user behavior data, effectively modeling long-term user behavior has become a research focus~\cite{youtube,MIMN}. While long-term behavior modeling offers insights into user preferences~\cite{SIM,ETA,SDIM,TWIN,TWIN2}. it also brings computational complexity, requiring a balance between model performance and online-service response time.

The industry typically adopts information filtering techniques to select the top K items, based on item attributes such as SIM~\cite{SIM}, or applying low-precision attention (e.g., ETA~\cite{ETA} and SDIM~\cite{SDIM}) to handle long sequences. This approach has been validated in various studies~\cite{URB4CTR}, including TWIN~\cite{TWIN} and TWIN v2~\cite{TWIN2}, which utilize GPU-optimized hierarchical attention to process sequences effectively, significantly enhancing CTR prediction accuracy. However, this approach sacrifices some items with low correlation. For example, when the target item is pants, the system focuses on pants-related behaviors for top K features but may overlook light-colored clothing purchases, which may imply an interest in dark pants.

To overcome these limitations and ensure efficient online computation, we explore alternative methodologies leveraging all data. Inspired by DeepSeek’s MLA~\cite{deepseekv2}, which compresses hidden representations for better performance and reduces KV cache storage. However, We find it unsuitable for our requirements as it does not reduce inference computation.

We propose Low-Rank Efficient Attention(LREA), a novel attention mechanism specifically designed for modeling long-term user behavior. LREA employs low-rank matrix decomposition at the sequence length level, optimizing performance while preserving the integrity of long-term behavior data. Furthermore, LREA integrates a tailor-designed loss function that maintains attention capabilities and simplifies the computation. In the inference phase, it utilizes matrix absorption and pre-storage techniques to effectively accommodate runtime constraints.

In summary, we propose LREA, the first approach focused on modeling entire long-term user behavior. LREA employs low-rank matrix decomposition and matrix absorption methods to meet online inference latency.

\section{Preliminaries}

\subsection{Task Formulation}
CTR prediction is formulated as a binary classification problem, wherein the objective is to learn a function \( f: \mathbb{R}^D \rightarrow \mathbb{R} \) based on a given training dataset \( \mathcal{D} = \{(x_1, y_1), \ldots, (x_{|\mathcal{D}|}, y_{|\mathcal{D}|})\} \). Each sparse feature of a training sample is converted into a low-dimensional dense representation through an embedding matrix. The embeddings generated from different features are concatenated as \( x_k \in \mathbb{R}^D \) and fed to a Multi-Layer Perceptron (MLP) layer, formalized as:
\begin{equation}
    \hat{y}_k = \sigma(MLP(x_k))
\end{equation}

where \( \sigma(\cdot) \) is the sigmoid function, and \( MLP \) is the mapping function that transforms the input \(x_k\) into a probability score. The optimization objective is the cross-entropy loss:
\begin{equation}
\mathcal{L} = -\frac{1}{| \mathcal{D} |} \sum_{k=1}^{| \mathcal{D} |} \left( y_k \log(\hat{y}_k) + (1 - y_k) \log(1 - \hat{y}_k) \right)
\end{equation}
where \( \mathcal{D} \) represents the training dataset, and \( y_k \in \{0, 1\} \) indicates the ground truth label for the corresponding interaction.

\subsection{long-term User Behavior Sequence}
 Behavior sequences provide a comprehensive view of users' historical interactions with various items. For each user \( u \in U \), we can model their behavioral history as a sequence of actions \( \mathcal{H}_{lu} = [s_1, s_2, \ldots, s_L] \), where \( s_{i} \) denotes the \( i \)-th interacted item. After being fed into the embedding layer, the representation of the \( i \)-th interacted item is \( e_i \in \mathbb{R }^{1 \times d}\) where d is the dimension of the representation of an item. Multi-head attention is widely applied to model user behavior sequence \cite{BST,AFM,DHAN,CAN}. In CTR prediction tasks, target item acts as query ($\boldsymbol{Q}$) and each item in user behavior sequence acts as key ($\boldsymbol{K}$) and value ($\boldsymbol{V}$). We call this multi-head attention structure as multi-head target attention, which is abbreviated as \textbf{\textit{MHTA}}. The online inference complexity of multi-head target attention is \( O(L*B*d) \), where \(B\) is the number of online inference candidate items and \( L\) is the length of the long-term user behavior sequence. As the length of the behavior sequence increases from tens to thousands, the Online computation latency is not tolerable.




\section{Low-Rank Efficient Attention}

\begin{figure*}[t]
  \centering
  \includegraphics[width=0.6\linewidth]{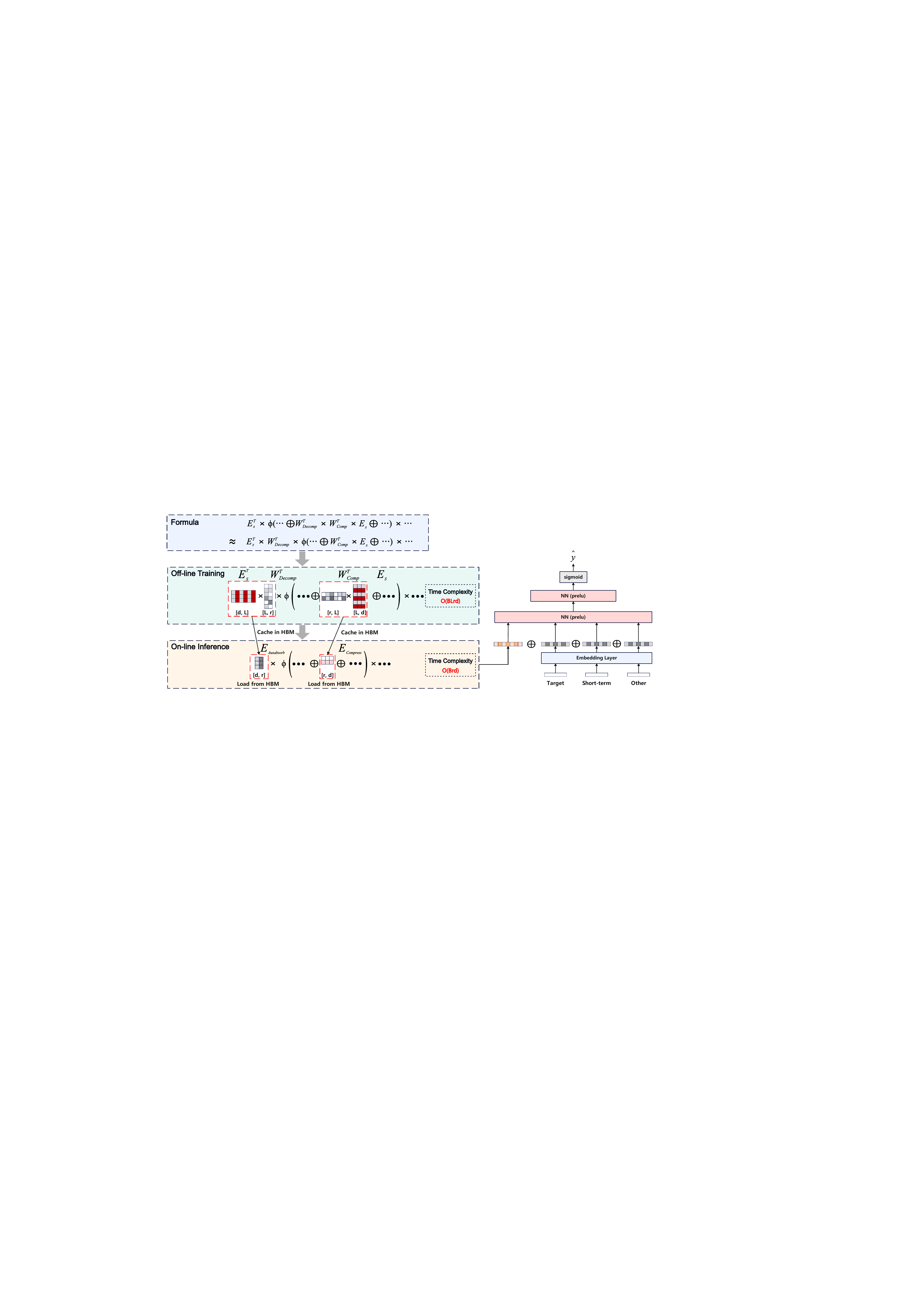}
  \vspace{-5pt}
  \caption{The structure of LREA. In the off-line training stage, LREA updates matrices \(W_{Comp}\) and \(W_{Decomp}\) without computational optimization, subsequently caching them with \(E_{s}\) and \(E_{s}^T\) in HBM. During offline inference, LREA utilizes compressed sequences \(E_{Auxabsorb}\) and \(E_{Comp}\), which reduce the length of user behavior sequence, enhancing the computation efficiency.}
  \Description{Diagram illustrating the LREA structure, showing training and inference stages with compressed sequences and matrices for efficient computation.}
  \label{fig:network}
  \vspace{-10pt}
\end{figure*}

In this section, we introduce LREA, a novel attention mechanism designed to tackle the inference limitations of modeling long-term user behavior sequences in CTR prediction. LREA applies low-rank matrix decomposition to user behavior sequences and uses matrix absorption techniques to address online runtime limitations.




\subsection{Long Sequence Compression}
CTR prediction fundamentally relies on the interaction of user behavior sequences and target items. Let \( \mathcal{H}_{lu} = [s_1, s_2, \ldots, s_L] \) denote a user's long-term behavior sequence, and \( E_t \in \mathbb{R}^d \) represent the embedding of the target item:

\begin{equation}
    TA(E_s, E_t) = E_s^T \cdot AttScore(E_s, E_t)
\end{equation}

In this equation, $E_s = [e_1, e_2, \ldots, e_L]^T$ is the embedding representation of behavior sequence and \( AttScore(E_s,\)
\(E_t) \in \mathbb{R}^{L\times1} \) represents the attention score designed to the behavior sequence.

\textbf{Low-rank Matrix Decompression} To mitigate information loss associated with pure compression, our approach, LREA, employs principles of low-rank matrix decomposition~\cite{lora}. Specifically, it utilizes two matrices \( W_{Comp} \) and \( W_{Decomp} \) to compress and subsequently decompress the original sequence representation:
\begin{equation}
    \label{eq:E_lrea}
    E_s^{LREA} = (E_s^T W_{Comp} W_{Decomp})^T
\end{equation}

Here, \( W_{Comp} \in \mathbb{R}^{L \times r} \) and \( W_{Decomp} \in \mathbb{R}^{r \times L} \), where \( L \) is the original sequence length and \( r \) is the reduced rank. \( E_{Comp}=E_s^T W_{Comp} \in \mathbb{R}^{d \times r} \) can be seen as a compressed representation of the behavior sequence. This operation helps to preserve essential information while handling the compressive aspect of the sequence. First we can formulate attention score computation of DIN~\cite{DIN} as:
\begin{equation}
    \label{eq:DIN}
    \begin{split}
        AttScore_{DIN}=&\phi \left[  \bigg( E_t  \oplus  E_s  \oplus  (E_t \odot E_s) \bigg) W_{in} \right]W_o \\
        =&\phi[ (E_tW_1)   + (E_s W_2) + (E_t \odot  E_s) W_3 ]W_o
    \end{split}
\end{equation}

where \(W_{in}, W_o\) are linear matrices and \(W_{in}\) could split into three parts \(W_1,W_2,W_3\) to calculate separately. \(\oplus\) means vector concatenation operation, \(\odot\) means hadamard production, \(\phi(\cdot)\) is a nonlinear activation function. Here we will emphasize the $(E_t \odot E_s)$ formulation's signification for interaction of the target item and sequence items' representation, and nonlinearity introduced by \(\phi(\cdot)\). These two components have the same function corresponding to the inner product \(QK^T\) and softmax in \textbf{\textit{MHTA}} respectively. Here we could introduce interaction more simply by $(E_t \odot  E_{Comp}^T)$, Then if we replace $E_s$ in DIN attention with $E_s^{LREA}$ in Eq. \ref{eq:E_lrea}, replace $(E_s \odot E_t)$ with $W_{Decomp}^T(E_t \odot  E_{Comp}^T)$, at the same time replace $E_t$ as $W_{Decomp}^TE_t$(after broadcast $E_t \in \mathbb{R}^{r\times d}$), we transform DIN attention score in formula \ref{eq:DIN} into LREA score $AttScore_{LREA}$:
\begin{footnotesize}
\begin{equation}
 \begin{split}
        &AttScore_{LREA}\\
        =&\phi[W_{Decomp}^TE_tW_1  \ + W_{Decomp}^TE_{Comp}^T W_2 + W_{Decomp}^T(E_t \odot  E_{Comp}^T) W_3]W_o \\
        =&\phi[W_{Decomp}^T(E_tW_1  \ + E_{Comp}^T W_2 + E_t \odot  E_{Comp}^T W_3)]W_o\\
    \end{split}
\end{equation}
\end{footnotesize}

\textbf{Matrix Absorption} If we can find a \(\phi(\cdot)\) satisfy:
\begin{equation}
    \label{eq:act}
    \phi(AB)\ = A\phi(B)
\end{equation}
we can easily derive this formula:
\begin{footnotesize}
\begin{equation}
\begin{split}
        LREA(E_s^T, E_t) &= E_s^T \cdot AttScore_{LREA} \\
        &= E_S^T W_{Decomp}^T \phi \bigg( [E_t \oplus E_{Comp}^T \oplus (E_t \odot E_{Comp}^T)]W_{in} \bigg)W_o
\end{split}
\end{equation}
\end{footnotesize}
The \(W_{Decomp}\) could be absorbed by \(E_{auxabsorb} = \bigg( E_S^T W_{Decomp}^T\bigg)_{(d \times r)}  \). just like \(E_{Comp}\), It can compress the L-dimension of \(E_s\) to a low-rank \(r\). Finally, the output of LREA will be concatenated with all other features' embedding and then fed into MLP to predict \(\hat{y}\).

A strict solution of $\phi(AB)\ = A\phi(B)$ is \( \phi(x) = kx, k\in\mathbb{R}\). But this will cause the network to lose nonlinearity.
A potential solution of nonlinear activation function \(\phi(\cdot)\) to approximate \( \phi(x) = kx\) is the ReLU function:
\begin{equation}
    ReLU(x) = \begin{cases} 
         x & \text{if } x \geq 0 \\
        0 & \text{if } x < 0 
        \end{cases}
\end{equation}

Because ReLU mostly meets the requirements of the solution to \(\phi(AB)\ = A\phi(B)\), and is more suitable than other activation functions. We could choose from the functions of ReLU series, Leaky ReLU, and Parametric ReLU\cite{prelu}. Here we use Leaky ReLU. No matter what variant ReLU function we choose, we need to guarantee to the greatest extent that the elements of matrix \(A\) and \(B\) have consistent signs, or both are non-negative. But if both \(A\) and \(B\) are non-negative, the formula is not nonlinear anymore. This will lead to sub-optimal performance for the attention mechanism. So there is a balance here between non-negativity and non-linearity.

\subsection{Model Training and Inference}
To ensure the validity of Eq. \ref{eq:act}, we need to impose non-negativity constraints on matrices \(A\) and \(B\). Therefore, we introduce a non-negativity loss function:
\begin{equation}
    \mathcal{L}_{non-neg} = || max(0, -W_{Decomp}^T)||_2^2 \ + \ || max(0, -W_{Comp}^TE_s)||_2^2
\end{equation}
where \(||A||_2 = \sqrt{\sum_{i,j} |A_{ij}}|^2 \) is the Frobenius norm of matrix A, so \(\mathcal{L}_{non-neg}\) will only impose a penalty on the negative elements in the matrix, while the penalty term for positive elements is 0.
The final training objective function is derived by:
\begin{equation}
    \mathcal{L}_{LREA} = \mathcal{L} \ + \lambda \ \mathcal{L}_{non-neg} 
\end{equation}

\begin{table}[]
\caption{Time complexity of different methods at training and online-serving stage.}
\vspace{-5pt}
\label{tab:complexity}
\begin{tabular}{@{}c|cc@{}}
\toprule
\textbf{Method} & \textbf{Training}       & \textbf{Online-Serving} \\ \midrule
DIN~\cite{DIN}             & O(BLd)                  & O(BLd)                  \\
SIM~\cite{SIM}             & O(Blog(M) + Bkd)        & O(Blog(M) + Bkd)        \\
ETA~\cite{ETA}             & O(Lmlog(d) + BLm + Bkd) & O(BLm + Bkd)            \\
SDIM~\cite{SDIM}            & O(Lmlog(d) + Bmlog(d))  & O(Bmlog(d))             \\
LREA            & O(BLrd)                 & O(Brd)                  \\ \bottomrule
\end{tabular}
\end{table}

Where \(\lambda\) is a hyper-parameter to balance the nonlinearity and non-negativity.

Fig.~\ref{fig:network} illustrates the overview of LREA. During the offline training stage, LREA trains and updates two additional matrices, \(W_{Compress}\) and \(W_{Decompress}\), without performing any optimization on computation. Once the training phase is complete, these two matrices are multiplied with \(E_{S}\) and \(E_{S}^T\), respectively, cached in the High Bandwidth Memory (HBM) denoted as \(E_{Compress}\) and \(E_{Auxabsorb}\).
In the online inference stage, LREA does not leverage the original long-term user behavior sequences. Instead, it utilizes the pre-stored sequences \(E_{Auxabsorb}\) and \(E_{Comp}\), in which the sequence length has been effectively reduced through absorption techniques. This method accelerates the inference process, ensuring a more efficient model deployment in a real-world system. By focusing on the compressed representations, LREA can deliver faster results while maintaining the integrity of the user behavior data, ultimately enhancing the overall performance of the CTR model.

Table~\ref{tab:complexity} shows the compute complexity of LREA compared with prior works.
In this analysis, the relevant parameters for our analysis include \(B\), the number of candidate items for each request; \(m\), the number of hashes employed; \(L\), \(k\) and \(r\), representing the lengths of the original, retrieved user behavior sequences and compressed user behavior sequences, respectively; \(M\), the size of the attribute inverted index in the SIM; and \(d\), the model's hidden size.

\section{Experiments}
In this section, we describe the experimental setup and conduct extensive experiments on public and industrial datasets to evaluate the performance and cost of the proposed \textbf{LERA} model.

\subsection{Experiments Setup}
\textbf{Datasets}. Our offline model evaluations are conducted on two widely used CTR prediction datasets and one industrial dataset.

\textbf{Amazon dataset\footnote{\url{https://cseweb.ucsd.edu/~jmcauley/datasets.html\#amazon_reviews}}}. It is derived from user interactions on Amazon’s e-commerce platform\cite{amz}. We use \textit{Books} collection of reviews, pre-process the data follows \cite{SIM}. The recent 10 user behaviors are short-term user sequential features and recent 90 user behaviors are long-term user sequential features.

\textbf{Taobao dataset\footnote{\url{https://tianchi.aliyun.com/dataset/dataDetail?dataId=649}}}. It is based on user behavior logs from Taobao, one of the major e-commerce platform in China and contains 89 million records. We take the click behaviors for each user and sort them according to time to construct the behavior sequence. Then we use the first 64 behaviors as short-term behavior sequential and the rest of 500 clicked items as long-term sequential features.

\textbf{Industrial dataset}. It is collected from the logs of our real-world industrial e-commerce advertising platform. The dataset encompasses advertising impression logs and the corresponding click labels during the period from December 3, 2024, to December 23, 2024. It contains 2 billion records from 10 million users. The user historical clicked items reach tens of thousands within the past year. We use the recent 64 click behaviors as short-term sequential features and 10000 items as long-term sequential features.

\begin{table}[t]
  \caption{Results on public datasets and industrial dataset}
  \label{tab:public and industrial result}
  \begin{tabular}{l|ll|ll}
    \toprule
    \hline
    \multirow{2}*{Models} & Amazon & Taobao &
    \multicolumn{2}{c}{Industrial dataset}\\ 
    \cline{2-5}
    & $AUC$ &$AUC$& $AUC$& $GAUC $\\
    \midrule
    DIN(Long sequence) & \textbf{0.8101}& 0.8903& 0.7149& 0.6337\\
    DIN(W/o long seq.)& 0.7833& 0.8535& 0.7005& 0.6187\\
    \midrule
    SIM & 0.7919& 0.8667& 0.7031& 0.6208\\
    ETA & 0.7984& 0.8651& 0.7083& 0.6235\\
    SDIM & 0.8013& 0.8734& 0.7097& 0.6239\\
    TWIN & 0.8015& 0.8838& 0.7103& 0.6258\\
    LREA & 0.8098& \textbf{0.8904}& \textbf{0.7154}& \textbf{0.6343}\\ 
    \hline
    \bottomrule
\end{tabular}
\end{table}

\textbf{Competitors}. We compare \textbf{LERA} with the mainstream CTR models as follows: {\bfseries DIN(W/o long seq.)}~\cite{DIN}: Only short-term sequences are involved. {\bfseries DIN(Long sequence)}: We apply the DIN target attention for both short and long-term sequences, but this method is not possible to deploy online due to intolerable inference latency. We also implement a two-stage method baseline for long-term user behaviors: {\bfseries SIM}~\cite{SIM}, {\bfseries ETA}~\cite{ETA}, {\bfseries SDIM}~\cite{SDIM}, {\bfseries TWIN}~\cite{TWIN}.

\textbf{Parameter Setting}. The hidden layer sizes are set to a fixed tuple [512, 256, 128]. Each hidden layer follows a Leaky ReLU \cite{lrelu} and a BatchNorm Layer \cite{BN} to accelerate the model training. All methods are trained with Adagrad for optimization. We tuned the learning rate in {0.01, 0.005, 0.001}. The batch size is set to 1024 and embedding dimension $d$ is set to 16.

\textbf{Evaluate Metrics}.
we use two widely adopted metrics: AUC and GAUC. GAUC performs a weighted average over all users' request group, and the weights are set to the number of samples of this user's request..We will only apply GAUC on industrial dataset due to the number of groups is relatively small in public datasets.

\begin{table}[]
    \caption{Hyper-parameters Analysis on Industrial Dataset}
    \label{tab:hyperParam}
    \centering
    \begin{tabular}{l|c|c|l|c|c}
        \toprule
        \hline
        \multicolumn{3}{c}{$\lambda=0.3$} & \multicolumn{3}{|c}{$r = 128$} \\
        \hline
        {LREA} & AUC & GAUC & {LREA} & AUC & GAUC \\
        \hline
        $r=32$ & 0.7081 & 0.6222 & $\lambda=0.05$ & 0.7121 & 06298 \\
        $r=64$ & 0.7097 & 0.6254 & $\lambda=0.1$ & 0.7138 & 06312 \\
        $r=128$ & \textbf{0.7154} & \textbf{0.6343} & $\lambda=0.3$ & \textbf{0.7154} & \textbf{0.6343} \\
         & - & - & $\lambda=0.5$ & +0.7137 & 0.6315 \\
         & - & - & $\lambda=0.8$ & +0.7129 & 0.6297 \\
        \hline
    \bottomrule
    \end{tabular}
\end{table}

\subsection{Experiment Performance}
{\bfseries Overall Perfomance.} The evaluation results is shown in Table \ref{tab:public and industrial result}. SIM, ETA, SDIM, TWIN, outperform DIN. That indicates long-term behavior sequence is helpful for CTR prediction since it can cover more user's historical interests. DIN(Long Sequence) has a stable improvement of AUC gain of 0.86\%/0.65\% compared with the best two-stage method on Amazon and Taobao dataset. We can also see a AUC gain of 0.49\% and a GAUC gain of 0.79\% in industrial dataset when comparing DIN(Long Sequence) with TWIN. But due to the limitation of inference latency, we can't deploy DIN to long sequences. However, our LREA also outperforms all two-stage methods and it achieves similar results to long sequence target attention, and even exceeds DIN on industrial datasets but its inference complexity is much less than that of DIN(Long sequence). 

{\bfseries Hyper-parameters Analysis.} We conduct analysis experiments of rank \(r\) and coefficient \(\lambda\) on our industrial dataset. As shown in Table \ref{tab:hyperParam}. First we fix \(\lambda\) to 0.3, tune r from 32 to 128. The rank of $r=128$ achieves the best performance, and AUC/GAUC drop significantly when $r$ becomes smaller. This indicates that we can't compression the representation of the original long sequences to a very low rank. We didn't test \(r\) > 128, because although the larger r is, the better performance we get, there should be a appropriate rank to balance cache memory size and model performance. Next, we fix \( r \) at 128 and adjust the value of \( \lambda \) between 0.05 and 0.8. As mentioned earlier, \( \lambda \) controls the nonlinearity and non-negativity in the compression module. If \( \lambda \) is too large, it reduces the nonlinearity, resulting in a decline in model performance. Conversely, if \( \lambda \) is too small, it skews the loss of the main task of CTR estimation. In our experiments, the model performs best when \( \lambda = 0.3 \).

{\bfseries Online A/B Test.} We conduct online A/B experiments to evaluate LREA in our online advertising system. The experiment lasts for 12 days and LREA achieves 5.88\% CTR and 4.26\% RPM gain compared to the product model (DIN+SIM). The inference rt only increases by 3 ms. And In comparison, the inference rt of DIN(Long sequence) will average increases about 30 ms, which causes 15\% of user requests to time out.
\section{Conclusion}
In this paper, we introduced LREA, a novel attention mechanism that enhances CTR prediction in online advertising and recommendation systems. 
LREA effectively integrates long-term user behavior by using low-rank matrix decomposition and matrix absorption. This mitigates the performance degradation caused by traditional top K selection. Experimental results demonstrate that our model achieves significant improvements over the competitors.

\bibliographystyle{ACM-Reference-Format}
\balance
\bibliography{ref}


\begin{thebibliography}{22}


\ifx \showCODEN    \undefined \def \showCODEN     #1{\unskip}     \fi
\ifx \showISBNx    \undefined \def \showISBNx     #1{\unskip}     \fi
\ifx \showISBNxiii \undefined \def \showISBNxiii  #1{\unskip}     \fi
\ifx \showISSN     \undefined \def \showISSN      #1{\unskip}     \fi
\ifx \showLCCN     \undefined \def \showLCCN      #1{\unskip}     \fi
\ifx \shownote     \undefined \def \shownote      #1{#1}          \fi
\ifx \showarticletitle \undefined \def \showarticletitle #1{#1}   \fi
\ifx \showURL      \undefined \def \showURL       {\relax}        \fi
\providecommand\bibfield[2]{#2}
\providecommand\bibinfo[2]{#2}
\providecommand\natexlab[1]{#1}
\providecommand\showeprint[2][]{arXiv:#2}

\bibitem[Cao et~al\mbox{.}(2022)]%
        {SDIM}
\bibfield{author}{\bibinfo{person}{Yue Cao}, \bibinfo{person}{Xiaojiang Zhou}, \bibinfo{person}{Jiaqi Feng}, \bibinfo{person}{Peihao Huang}, \bibinfo{person}{Yao Xiao}, \bibinfo{person}{Dayao Chen}, {and} \bibinfo{person}{Sheng Chen}.} \bibinfo{year}{2022}\natexlab{}.
\newblock \showarticletitle{Sampling is all you need on modeling long-term user behaviors for CTR prediction}. In \bibinfo{booktitle}{\emph{Proceedings of the 31st ACM International Conference on Information \& Knowledge Management}}. \bibinfo{pages}{2974--2983}.
\newblock


\bibitem[Chang et~al\mbox{.}(2023)]%
        {TWIN}
\bibfield{author}{\bibinfo{person}{Jianxin Chang}, \bibinfo{person}{Chenbin Zhang}, \bibinfo{person}{Zhiyi Fu}, \bibinfo{person}{Xiaoxue Zang}, \bibinfo{person}{Lin Guan}, \bibinfo{person}{Jing Lu}, \bibinfo{person}{Yiqun Hui}, \bibinfo{person}{Dewei Leng}, \bibinfo{person}{Yanan Niu}, \bibinfo{person}{Yang Song}, {et~al\mbox{.}}} \bibinfo{year}{2023}\natexlab{}.
\newblock \showarticletitle{TWIN: TWo-stage interest network for lifelong user behavior modeling in CTR prediction at kuaishou}. In \bibinfo{booktitle}{\emph{Proceedings of the 29th ACM SIGKDD Conference on Knowledge Discovery and Data Mining}}. \bibinfo{pages}{3785--3794}.
\newblock


\bibitem[Chen et~al\mbox{.}(2021)]%
        {ETA}
\bibfield{author}{\bibinfo{person}{Qiwei Chen}, \bibinfo{person}{Changhua Pei}, \bibinfo{person}{Shanshan Lv}, \bibinfo{person}{Chao Li}, \bibinfo{person}{Junfeng Ge}, {and} \bibinfo{person}{Wenwu Ou}.} \bibinfo{year}{2021}\natexlab{}.
\newblock \showarticletitle{End-to-end user behavior retrieval in click-through rateprediction model}.
\newblock \bibinfo{journal}{\emph{arXiv preprint arXiv:2108.04468}} (\bibinfo{year}{2021}).
\newblock


\bibitem[Chen et~al\mbox{.}(2019)]%
        {BST}
\bibfield{author}{\bibinfo{person}{Qiwei Chen}, \bibinfo{person}{Huan Zhao}, \bibinfo{person}{Wei Li}, \bibinfo{person}{Pipei Huang}, {and} \bibinfo{person}{Wenwu Ou}.} \bibinfo{year}{2019}\natexlab{}.
\newblock \showarticletitle{Behavior sequence transformer for e-commerce recommendation in Alibaba}. In \bibinfo{booktitle}{\emph{Proceedings of the 1st International Workshop on Deep Learning Practice for High-Dimensional Sparse Data}} (Anchorage, Alaska) \emph{(\bibinfo{series}{DLP-KDD '19})}. Article \bibinfo{articleno}{12}, \bibinfo{numpages}{4}~pages.
\newblock


\bibitem[Covington et~al\mbox{.}(2016)]%
        {youtube}
\bibfield{author}{\bibinfo{person}{Paul Covington}, \bibinfo{person}{Jay Adams}, {and} \bibinfo{person}{Emre Sargin}.} \bibinfo{year}{2016}\natexlab{}.
\newblock \showarticletitle{Deep Neural Networks for YouTube Recommendations}. In \bibinfo{booktitle}{\emph{Proceedings of the 10th ACM Conference on Recommender Systems}} (Boston, Massachusetts, USA) \emph{(\bibinfo{series}{RecSys '16})}. \bibinfo{pages}{191–198}.
\newblock


\bibitem[DeepSeek-AI et~al\mbox{.}(2024)]%
        {deepseekv2}
\bibfield{author}{\bibinfo{person}{DeepSeek-AI}, \bibinfo{person}{Aixin Liu}, \bibinfo{person}{Bei Feng}, \bibinfo{person}{Bin Wang}, \bibinfo{person}{Bingxuan Wang}, {et~al\mbox{.}}} \bibinfo{year}{2024}\natexlab{}.
\newblock \bibinfo{title}{DeepSeek-V2: A Strong, Economical, and Efficient Mixture-of-Experts Language Model}.
\newblock
\showeprint[arxiv]{2405.04434}~[cs.CL]
\urldef\tempurl%
\url{https://arxiv.org/abs/2405.04434}
\showURL{%
\tempurl}


\bibitem[He et~al\mbox{.}(2015)]%
        {prelu}
\bibfield{author}{\bibinfo{person}{Kaiming He}, \bibinfo{person}{Xiangyu Zhang}, \bibinfo{person}{Shaoqing Ren}, {and} \bibinfo{person}{Jian Sun}.} \bibinfo{year}{2015}\natexlab{}.
\newblock \showarticletitle{Delving Deep into Rectifiers: Surpassing Human-Level Performance on ImageNet Classification}. In \bibinfo{booktitle}{\emph{Proceedings of the 2015 IEEE International Conference on Computer Vision (ICCV)}} \emph{(\bibinfo{series}{ICCV '15})}. \bibinfo{pages}{1026–1034}.
\newblock


\bibitem[Hu et~al\mbox{.}(2022)]%
        {lora}
\bibfield{author}{\bibinfo{person}{Edward~J Hu}, \bibinfo{person}{yelong shen}, \bibinfo{person}{Phillip Wallis}, \bibinfo{person}{Zeyuan Allen-Zhu}, \bibinfo{person}{Yuanzhi Li}, \bibinfo{person}{Shean Wang}, \bibinfo{person}{Lu Wang}, {and} \bibinfo{person}{Weizhu Chen}.} \bibinfo{year}{2022}\natexlab{}.
\newblock \showarticletitle{Lo{RA}: Low-Rank Adaptation of Large Language Models}. In \bibinfo{booktitle}{\emph{International Conference on Learning Representations}} \emph{(\bibinfo{series}{ICLR '22})}.
\newblock


\bibitem[Ioffe(2015)]%
        {BN}
\bibfield{author}{\bibinfo{person}{Sergey Ioffe}.} \bibinfo{year}{2015}\natexlab{}.
\newblock \showarticletitle{Batch normalization: Accelerating deep network training by reducing internal covariate shift}.
\newblock \bibinfo{journal}{\emph{arXiv preprint arXiv:1502.03167}} (\bibinfo{year}{2015}).
\newblock


\bibitem[Lou et~al\mbox{.}(2024)]%
        {UID}
\bibfield{author}{\bibinfo{person}{Jiazhen Lou}, \bibinfo{person}{Zhao Li}, \bibinfo{person}{Hong Wen}, \bibinfo{person}{Jingsong Lv}, \bibinfo{person}{Jing Zhang}, \bibinfo{person}{Fuyu Lv}, \bibinfo{person}{Zulong Chen}, {and} \bibinfo{person}{Jia Wu}.} \bibinfo{year}{2024}\natexlab{}.
\newblock \showarticletitle{UID-Net: Enhancing Click-Through Rate Prediction in Trigger-Induced Recommendation Through User Interest Decomposition}. In \bibinfo{booktitle}{\emph{Advanced Data Mining and Applications: 20th International Conference, ADMA 2024, Sydney, NSW, Australia, December 3–5, 2024, Proceedings, Part VI}}. \bibinfo{pages}{49–64}.
\newblock


\bibitem[Maas et~al\mbox{.}(2013)]%
        {lrelu}
\bibfield{author}{\bibinfo{person}{Andrew~L Maas}, \bibinfo{person}{Awni~Y Hannun}, \bibinfo{person}{Andrew~Y Ng}, {et~al\mbox{.}}} \bibinfo{year}{2013}\natexlab{}.
\newblock \showarticletitle{Rectifier nonlinearities improve neural network acoustic models}. In \bibinfo{booktitle}{\emph{Proc. icml}}, Vol.~\bibinfo{volume}{30}. Atlanta, GA, \bibinfo{pages}{3}.
\newblock


\bibitem[Ni et~al\mbox{.}(2019)]%
        {amz}
\bibfield{author}{\bibinfo{person}{Jianmo Ni}, \bibinfo{person}{Jiacheng Li}, {and} \bibinfo{person}{Julian McAuley}.} \bibinfo{year}{2019}\natexlab{}.
\newblock \showarticletitle{Justifying recommendations using distantly-labeled reviews and fine-grained aspects}. In \bibinfo{booktitle}{\emph{Proceedings of the 2019 conference on empirical methods in natural language processing and the 9th international joint conference on natural language processing (EMNLP-IJCNLP)}}. \bibinfo{pages}{188--197}.
\newblock


\bibitem[Pi et~al\mbox{.}(2019)]%
        {MIMN}
\bibfield{author}{\bibinfo{person}{Qi Pi}, \bibinfo{person}{Weijie Bian}, \bibinfo{person}{Guorui Zhou}, \bibinfo{person}{Xiaoqiang Zhu}, {and} \bibinfo{person}{Kun Gai}.} \bibinfo{year}{2019}\natexlab{}.
\newblock \showarticletitle{Practice on Long Sequential User Behavior Modeling for Click-Through Rate Prediction}. In \bibinfo{booktitle}{\emph{Proceedings of the 25th ACM SIGKDD International Conference on Knowledge Discovery \& Data Mining}} (Anchorage, AK, USA) \emph{(\bibinfo{series}{KDD '19})}. \bibinfo{pages}{2671–2679}.
\newblock


\bibitem[Pi et~al\mbox{.}(2020)]%
        {SIM}
\bibfield{author}{\bibinfo{person}{Qi Pi}, \bibinfo{person}{Guorui Zhou}, \bibinfo{person}{Yujing Zhang}, \bibinfo{person}{Zhe Wang}, \bibinfo{person}{Lejian Ren}, \bibinfo{person}{Ying Fan}, \bibinfo{person}{Xiaoqiang Zhu}, {and} \bibinfo{person}{Kun Gai}.} \bibinfo{year}{2020}\natexlab{}.
\newblock \showarticletitle{Search-based user interest modeling with lifelong sequential behavior data for click-through rate prediction}. In \bibinfo{booktitle}{\emph{Proceedings of the 29th ACM International Conference on Information \& Knowledge Management}}. \bibinfo{pages}{2685--2692}.
\newblock


\bibitem[Qin et~al\mbox{.}(2020)]%
        {URB4CTR}
\bibfield{author}{\bibinfo{person}{Jiarui Qin}, \bibinfo{person}{Weinan Zhang}, \bibinfo{person}{Xin Wu}, \bibinfo{person}{Jiarui Jin}, \bibinfo{person}{Yuchen Fang}, {and} \bibinfo{person}{Yong Yu}.} \bibinfo{year}{2020}\natexlab{}.
\newblock \showarticletitle{User behavior retrieval for click-through rate prediction}. In \bibinfo{booktitle}{\emph{Proceedings of the 43rd International ACM SIGIR Conference on Research and Development in Information Retrieval}}. \bibinfo{pages}{2347--2356}.
\newblock


\bibitem[Shen et~al\mbox{.}(2022)]%
        {DIHN}
\bibfield{author}{\bibinfo{person}{Qijie Shen}, \bibinfo{person}{Hong Wen}, \bibinfo{person}{Wanjie Tao}, \bibinfo{person}{Jing Zhang}, \bibinfo{person}{Fuyu Lv}, \bibinfo{person}{Zulong Chen}, {and} \bibinfo{person}{Zhao Li}.} \bibinfo{year}{2022}\natexlab{}.
\newblock \showarticletitle{Deep interest highlight network for click-through rate prediction in trigger-induced recommendation}. In \bibinfo{booktitle}{\emph{Proceedings of the ACM web conference 2022}}. \bibinfo{pages}{422--430}.
\newblock


\bibitem[Si et~al\mbox{.}(2024)]%
        {TWIN2}
\bibfield{author}{\bibinfo{person}{Zihua Si}, \bibinfo{person}{Lin Guan}, \bibinfo{person}{ZhongXiang Sun}, \bibinfo{person}{Xiaoxue Zang}, \bibinfo{person}{Jing Lu}, \bibinfo{person}{Yiqun Hui}, \bibinfo{person}{Xingchao Cao}, \bibinfo{person}{Zeyu Yang}, \bibinfo{person}{Yichen Zheng}, \bibinfo{person}{Dewei Leng}, {et~al\mbox{.}}} \bibinfo{year}{2024}\natexlab{}.
\newblock \showarticletitle{Twin v2: Scaling ultra-long user behavior sequence modeling for enhanced ctr prediction at kuaishou}. In \bibinfo{booktitle}{\emph{Proceedings of the 33rd ACM International Conference on Information and Knowledge Management}}. \bibinfo{pages}{4890--4897}.
\newblock


\bibitem[Xiao et~al\mbox{.}(2017)]%
        {AFM}
\bibfield{author}{\bibinfo{person}{Jun Xiao}, \bibinfo{person}{Hao Ye}, \bibinfo{person}{Xiangnan He}, \bibinfo{person}{Hanwang Zhang}, \bibinfo{person}{Fei Wu}, {and} \bibinfo{person}{Tat-Seng Chua}.} \bibinfo{year}{2017}\natexlab{}.
\newblock \showarticletitle{Attentional factorization machines: learning the weight of feature interactions via attention networks}. In \bibinfo{booktitle}{\emph{Proceedings of the 26th International Joint Conference on Artificial Intelligence}} (Melbourne, Australia) \emph{(\bibinfo{series}{IJCAI'17})}. \bibinfo{pages}{3119–3125}.
\newblock


\bibitem[Xu et~al\mbox{.}(2020)]%
        {DHAN}
\bibfield{author}{\bibinfo{person}{Weinan Xu}, \bibinfo{person}{Hengxu He}, \bibinfo{person}{Minshi Tan}, \bibinfo{person}{Yunming Li}, \bibinfo{person}{Jun Lang}, {and} \bibinfo{person}{Dongbai Guo}.} \bibinfo{year}{2020}\natexlab{}.
\newblock \showarticletitle{Deep Interest with Hierarchical Attention Network for Click-Through Rate Prediction}. In \bibinfo{booktitle}{\emph{Proceedings of the 43rd International ACM SIGIR Conference on Research and Development in Information Retrieval}} (Virtual Event, China) \emph{(\bibinfo{series}{SIGIR '20})}. \bibinfo{pages}{1905–1908}.
\newblock


\bibitem[Zhang et~al\mbox{.}(2024)]%
        {CAN}
\bibfield{author}{\bibinfo{person}{Hengyu Zhang}, \bibinfo{person}{Junwei Pan}, \bibinfo{person}{Dapeng Liu}, \bibinfo{person}{Jie Jiang}, {and} \bibinfo{person}{Xiu Li}.} \bibinfo{year}{2024}\natexlab{}.
\newblock \showarticletitle{Deep Pattern Network for Click-Through Rate Prediction}. In \bibinfo{booktitle}{\emph{Proceedings of the 47th International ACM SIGIR Conference on Research and Development in Information Retrieval}} (Washington DC, USA) \emph{(\bibinfo{series}{SIGIR '24})}. \bibinfo{pages}{1189–1199}.
\newblock


\bibitem[Zhou et~al\mbox{.}(2019)]%
        {DIEN}
\bibfield{author}{\bibinfo{person}{Guorui Zhou}, \bibinfo{person}{Na Mou}, \bibinfo{person}{Ying Fan}, \bibinfo{person}{Qi Pi}, \bibinfo{person}{Weijie Bian}, \bibinfo{person}{Chang Zhou}, \bibinfo{person}{Xiaoqiang Zhu}, {and} \bibinfo{person}{Kun Gai}.} \bibinfo{year}{2019}\natexlab{}.
\newblock \showarticletitle{Deep interest evolution network for click-through rate prediction}. In \bibinfo{booktitle}{\emph{Proceedings of the Thirty-Third AAAI Conference on Artificial Intelligence and Thirty-First Innovative Applications of Artificial Intelligence Conference and Ninth AAAI Symposium on Educational Advances in Artificial Intelligence}} (Honolulu, Hawaii, USA) \emph{(\bibinfo{series}{AAAI'19/IAAI'19/EAAI'19})}. Article \bibinfo{articleno}{729}, \bibinfo{numpages}{8}~pages.
\newblock


\bibitem[Zhou et~al\mbox{.}(2018)]%
        {DIN}
\bibfield{author}{\bibinfo{person}{Guorui Zhou}, \bibinfo{person}{Xiaoqiang Zhu}, \bibinfo{person}{Chenru Song}, \bibinfo{person}{Ying Fan}, \bibinfo{person}{Han Zhu}, \bibinfo{person}{Xiao Ma}, \bibinfo{person}{Yanghui Yan}, \bibinfo{person}{Junqi Jin}, \bibinfo{person}{Han Li}, {and} \bibinfo{person}{Kun Gai}.} \bibinfo{year}{2018}\natexlab{}.
\newblock \showarticletitle{Deep interest network for click-through rate prediction}. In \bibinfo{booktitle}{\emph{Proceedings of the 24th ACM SIGKDD international conference on knowledge discovery \& data mining}}. \bibinfo{pages}{1059--1068}.
\newblock


\end{thebibliography}

\end{document}